\documentclass[aps,prd,twocolumn,groupedaddress,showpacs]{revtex4}

\usepackage{graphicx}
\usepackage{graphicx}
\usepackage{amssymb}
\usepackage{amsmath}
\usepackage{color}
\newcommand{\be}{\begin{equation}}
\newcommand{\ee}{\end{equation}}
 
 \definecolor{BrickRed}{cmyk}{0,0.89,0.94,0.28}
\definecolor{MidnightBlue}{cmyk}{0.98,0.13,0,0.43}
\definecolor{DarkGreen}{rgb}{0,0.7,0.1}

\begin{document}

\title{Classical Casimir interaction of perfectly conducting sphere and plate.}


\author{Giuseppe Bimonte}

\affiliation{ Dipartimento di Fisica E. Pancini, Universit\`{a} di
Napoli Federico II, Complesso Universitario
di Monte S. Angelo,  Via Cintia, I-80126 Napoli, Italy}
\affiliation{INFN Sezione di Napoli, I-80126 Napoli, Italy}

\email{giuseppe.bimonte@na.infn.it}

\begin{abstract}

We study the Casimir interaction between perfectly conducting sphere and plate in the classical limit of high temperatures. By taking the small-distance expansion of the exact
scattering formula, we compute the leading correction to the Casimir energy beyond the commonly employed proximity force approximation. We find that for a sphere of radius $R$ at distance $d$ from the plate the correction is of the form $\ln^2 (d/R)$, in agreement with indications from recent large-scale numerical  computations. We develop a fast-converging numerical scheme for computing the Casimir interaction to high precision, based on   bispherical partial waves, and we verify that the short-distance formula  provides precise values of the Casimir energy  also for fairly large distances.

\end{abstract}


\maketitle

\section{Introduction}
\label{sec:intro}

Over the last two decades a new wave of precision experiments   spurred  a strong resurgence of interest in the Casimir effect \cite{Casimir48}, the tiny long-range force between (neutral) macroscopic polarizable bodies, due to quantum and thermal charge fluctuations within the bodies. For  reviews  we address the reader  to Refs.\cite{book1,parse,book2,woods}.

A distinctive feature of the Casimir force is its {\it non-additivity},   a consequence of the inherent many-body character of fluctuation forces. This property  enormously complicates the task of computing the Casimir force in arbitrary geometries. Indeed, Casimir himself was able to compute the force only for the highly idealized geometry of two perfectly conducting plane-parallel surfaces at zero temperature.
A major step forward was made a few years later  by Lifshitz \cite{lifs}, who derived an exact formula for the Casimir interaction between two plane-parallel dielectric slabs at finite temperature.
After these early successes in the planar geometry, the task of computing the force in non-planar geometries remained  untractable  for decades. This limitation has  represented a serious practical problem because, in order to avoid the insurmountable parallelism issues posed by two plane-parallel  surfaces separated by a sub-micron separation,  practically all Casimir experiments adopt non planar geometries, like the sphere-plate. For a long time, the only tool to estimate the Casimir force  between two non-planar gently curved surfaces has been the old-fashioned Proximity Force Approximation (PFA) \cite{Derjaguin}, which expresses the Casimir force as the average of the plane-parallel force  over the local separation between the surfaces.  Being an uncontrollable approximation, the problem remained  of addressing the systematic theoretical  error engendered  by the PFA, an issue of  great importance  for a proper interpretation of modern   high precision  Casimir experiments.

A breakthrough occurred in the early 2000's when, generalizing   early findings by Balian and Duplantier \cite{Balian}  and Langbein \cite{langbein}, an exact scattering formula for the Casimir interaction between two (or more) dielectric objects of any shape was finally worked out  \cite{sca1,sca2, kenneth}. Shortly later, scattering formulae  have been derived for the Casimir force  and  the power of heat transfer between two bodies at different temperatures \cite{bimontescat,antezza1,krugertrace}. At thermal equilibrium, the scattering formula has the form of a sum over so-called Matsubara (imaginary) frequencies of functional determinants involving the  T-operators   of the involved bodies. In principle, the formula allows to exactly  compute the Casimir interaction between two bodies whose  T-operators are either known (as it is the case for planes, spheres or cylinders \cite{emiguniversal}) or can be computed numerically  (for example for periodic rectangular  gratings  \cite{lambrgrat}).   Numerical schemes inspired by the scattering formalism have been developed over the last few years, by which it is now possible to estimate numerically with reasonable precision the Casimir force between objects with complicated shapes (see \cite{johnson} and Refs. therein). Very recently,  the scattering formula  has been  evaluated to yield the exact classical Casimir interaction between a sphere and a plate subjected to Drude (Dr) boundary conditions \cite{bimonteex1},  and between two perfectly conducting spheres in four Euclidean dimensions \cite{bimonteex2}.

The   scattering formula undoubtedly constitutes a major advancement, however  its practical utility has been limited so far  because there is no known method to compute efficiently  the  T-operator for material surfaces of arbitrary shapes.  Even in the few cases  (spheres or cylinders) for which the scattering operator is known exactly, use of the scattering formula is hampered by its  slow convergence rate.  Consider as an example the experimentally important geometry of a metallic sphere of radius $R$ at a minimum distance $d$ from a plane. It has been found that to accurately compute the Casimir interaction, it is  necessary to    include in the scattering formula all multipoles   up to an order   $ l_{\rm max} \gtrsim R/d$ or so. At the moment of this writing, the
largest numerical computation  \cite{lambrsphere}    has  $l_{\rm max}=100$, which allows to estimate the Casimir interaction only for aspects ratios $d/R$  smaller than  0.02 or so. For comparison, it should be considered that  in order to increase the magnitude of the force, Casimir experiments   use a large sphere   at small distances from the plate, with typical aspects ratios   of the order of one thousandth. For so small aspect ratios, a precise computation of the Casimir interaction requires  multipole orders of several thousands, which are out of reach for now.

From a practical standpoint, the main  role of the exact scattering formula has perhaps been to  serve as a guide towards systematically deriving  approximation schemes  in various regimes, going beyond the old PFA. For surfaces carrying corrugations of small amplitude, a systematic perturbative expansion of the Casimir interaction in powers of the small corrugation amplitude has been worked out \cite{lambrcorr}.  Several researchers have instead   endeavored to compute   curvature corrections to the Casimir interaction, in the experimentally important limit of small surface separations. This is clearly a problem of outmost practical importance, for the purpose of interpreting current small-distance precision experiments. There are presently two approaches to compute curvature corrections to the PFA. The first one consists in working out the asymptotic small-distance expansion of the exact scattering formula.  The method is rigorous, but it has the drawback   that the expansion has to be worked out {\it ab initio} for each model, and for each surface geometry.   By following this route, the next-to-leading-order (NTLO) correction to the Casimir energy has been computed for the cylinder/plate and the sphere/plate geometries, initially for a free scalar field obeying Dirichlet (D) boundary conditions (bc) \cite{bordag1}, and then for the electromagnetic (em) field with perfect-conductor (P) bc \cite{bordag2}. Later  the same approach was applied to  a free scalar field obeying  D, Neumann (N) and mixed ND bc on two parallel cylinders \cite{teo1}.

An alternative route to compute  the NTLO correction to PFA    assumes that   the Casimir energy functional admits   a derivative expansion (DE)  in   powers of derivatives of the surfaces height profiles. The coefficients of the DE are computed by matching the DE with the perturbative expansion of the Casimir-energy functional in the common domain of validity (for details see \cite{fosco1,bimonte1}).  An advantage offered by the DE, in comparison with the previous approach, is that once the DE  is worked out for a specific model, it can be straightforwardly applied to  surfaces of any shape.  In \cite{fosco1} the DE  was  worked out for a D scalar field in the cylinder and sphere/plate geometries,  giving results in agreement with the asymptotic small-distance expansion of the scattering formula in \cite{bordag1}. The DE  for the em field with  P  bc, as well as for a scalar field obeying N and mixed DN bc was later  worked out in \cite{bimonte1}, where the DE was also generalized  to the  case of two curved surfaces.   Interestingly, the  NTLO correction for  the sphere/plate geometry with P bc obtained in \cite{bimonte1}  by using the DE   was in disagreement with the result  reported in \cite{bordag2}: while the DE predicted an analytic  correction $\sim d/R$,   a larger logarithmic $\sim d/R\,\ln(d/R)$ correction had been found in \cite{bordag2}.
A successive recalculation by some of the authors of \cite{bordag2} detected a sign mistake in their original computation, and finally  led to full agreement  with the DE expansion also in em  and N cases.  The DE for a D and N scalar at zero and finite temperature in any number of space-time dimensions   was worked out in \cite{fosco2}, while the experimentally important case of dielectric curved surfaces at finite temperature  is presented in \cite{bimonte2}. It is worth stressing that the NTLO correction predicted by the DE is also in full agreement with the  short distance expansion of the  exact sphere-plate and sphere-sphere
classical Casimir energies both for Dr bc \cite{bimonteex1}  as well as for P bc in four Euclidean dimensions  \cite{bimonteex2}. The DE has been  also used to study curvature effects in
the Casimir-Polder interaction of a particle with a gently curved surface \cite{CPbimonte1,CPbimonte2}. The same method has been used very recently to estimate the shifts of the rotational
levels of a diatomic molecule due to its van der Waals interaction with a curved dielectric surface \cite{CPbimonte3}.

In this paper we study the    sphere-plate Casimir interaction for P bc, in the   high-temperature (HT) or classical limit. In this limit, the Casimir interaction  reduces to the zero-frequency Matsubara-term of the full   finite-temperature scattering formula.
The zero-frequency (i.e. the classical) term becomes dominant  for sphere-plate separations $d$ that are larger than the thermal length $\lambda_T=\hbar c/(k_B T)$ ($\lambda_T=7.5$ microns at room temperature). A perfect conductor constitutes the idealized limit of a superconductor, i.e. a conductor with perfect Meissner effect \cite{bimontesuper}. Ohmic metals  are better modeled as Drude conductors, since normal metals do not impede static magnetic fields.  Quite surprisingly, several short-distance precision experiments (see \cite{deccamag} and Refs. therein) with metallic plates at room temperature are in better agreement with a  superconductor-like model (i.e. the plasma model) for the dielectric function of the plates, while a single large distance experiment \cite{lamorth} favors the Drude model.  For a thorough discussion of this delicate problem we address the reader to the monograph \cite{book2}.

The HT limit of the Casimir interaction for P bc has been investigated in \cite{irina}, where the asymptotic small distance expansion of the scattering formula was shown to reproduce in leading order the PFA. The authors of \cite{irina} did not study though corrections to PFA. Determining the form of the the NTLO correction  
is an interesting problem, for the following reason. In the HT limit, the Casimir interaction  for P bc is mathematically equivalent to the sum of the classical Casimir energies for a  Dr or D scalar field (depending on whether the plates are grounded or not) plus a N scalar field. The HT limit of the sphere-plate Casimir interaction for D and Dr bc have been computed exactly not long ago \cite{bimonteex1}. However, the N and  P cases have been intractable so far. Working out the NTLO correction to PFA for these two models is of great interest, because in the HT limit  the perturbative kernels for  N and P bc display a singular behavior for small in-plane momenta, invalidating  the DE  \cite{fosco2,bimonteex1} \footnote{The DE exists however for N and  P bc at zero temperature \cite{bimonte1}.}. The DE has been shown to fail also for the plasma model in the HT limit in \cite{foscopl}.  As a result, the analytic form of the NTLO for N and  P bc is so far unknown.  A large-scale  numerical computation including up to 5000 partial waves \cite{antoine} suggests a $\ln^2 (d/R)$ form for the NTLO term.  However, the data of  \cite{antoine}  appeared to support a $\ln^2 (d/R)$  also for the Dr model, and from the exact solution in \cite{bimonteex1} we now know that the correct  NTLO correction for the Dr model  is actually  a $\ln (d/R)$ term, in accordance with the DE.  To resolve the matter, it is clearly of interest to see if the NTLO for  N and  P bc can be worked out analytically. As we said above the DE cannot help now, and therefore we attacked the problem using the method based on the asymptotic expansion of the scattering formula \cite{bordag1,teo1}.  We find that the NTLO is indeed of the $\ln^2 (d/R)$ form as  it was argued  in \cite{antoine}, and we determine its coefficient. We also develop an efficient numerical implementation of the exact scattering formula, based on the use of bispherical multipoles \cite{bimonteex1}. The fast convergence of our scheme allowed us to probe   extremely small aspect ratios down to $d/R=10^{-5}$. We verify that the approximate formula obtained  by taking the asymptotic short-distance expansion  of the Casimir interaction is actually very accurate up to relatively large values of the aspect ratio.

The paper is organized as follows:  in Sec II we discuss the HT limit of the scattering formula, and briefly review the exact HT sphere-plate solution for D and Dr bc  of \cite{bimonteex1}. In Sec. III we compute  the  short-distance expansion of the HT scattering formula in the sphere-plate geometry for P bc, and we obtain an approximate formula for the Casimir interaction valid for small separations. By taking its short-distance limit, we compute explicitly the leading correction of the Casimir interaction beyond PFA.  In Sec. IV we present a fast-convergent numerical scheme to compute the Casimir energy  based on the use of bispherical multipoles, and compare our numerical results with the approximate formula derived in Sec. III. In Sec. V we present our conclusions.

\section{The Casimir energy in the classical limit}

We start from the general scattering formula \cite{sca1,sca2,kenneth} for the Casimir free energy  of two objects (denoted as 1 and 2) in vacuum:
$$
{\cal F}=k_B T \sum_{n \ge 0}\,\!\!' \,{\rm Tr \ln} [1-{\hat M}({\rm i} \,\xi_n)]\;,
$$
\be
{\hat M}={\hat T}^{(1)}{\hat U}{\hat T}^{(2)}{\hat U}\;.\label{scafor}
\ee
Here $k_B$ is Boltzmann's constant, $T$ is the temperature,   $\xi_n= 2 \pi n k_B T/\hbar$ are the (imaginary) Matsubara frequencies, and the prime in the sum indicates that the $n=0$ term is taken with weight 1/2.  In Eq. (\ref{scafor}), ${\hat T}^{(j)}$ denotes the T-operator of object $j$, evaluated for imaginary frequency ${\rm i} \,\xi_n$, and ${\hat U}$ is the translation operator that translates the scattering solution from the coordinate of one object to the one of the other object. When considered in a plane-wave basis $|{\bf k},Q \rangle$, where ${\bf k}$ is the in-plane wave-vector and $Q=E,M$ is the polarization ($E$ and $M$ denote, respectively, transverse magnetic and transverse electric modes), the translation operator ${\hat U}$ is diagonal, with matrix elements $e^{-d q_n}$ where $d$ is the minimum distance between the objects,  $q_n=\sqrt{k^2+\xi_n^2/c^2}$ with $k=|{\bf k}|$, and $c$ the speed of light. This shows that in the  HT limit $k_B T \gg \hbar c/d$, the free energy is dominated by the first term $n=0$ in the sum Eq. (\ref{scafor}):
\be
{\cal F}_{\rm HT}=-k_B T \;\Phi,\;\;\;\;\Phi=-\frac{1}{2} {\rm Tr \ln} [1-{\hat M}(0)]\;.\label{HTscat}
\ee
Here, $\Phi$ is a dimensionless temperature-independent function, depending on the {\it static} em response functions of the two bodies. Since the free-energy is proportional to the temperature,  the HT (or classical) limit of the Casimir interaction has a purely entropic character.

We are interested in the classical Casimir interaction ${\cal F}_{\rm HT}$ of a   sphere of radius $R$ placed at a (minimum) distance $d$ from a  plate, bot subjected to P bc. We take the surface of the plate to coincide with the $(x,y)$ plane of a cartesian coordinate system, whose $z$ axis passes through the sphere center $C$  (see Fig.1). We define the aspect ratio $x$ of the system as $x=d/R$.
\begin{figure}
\includegraphics [width=.9\columnwidth]{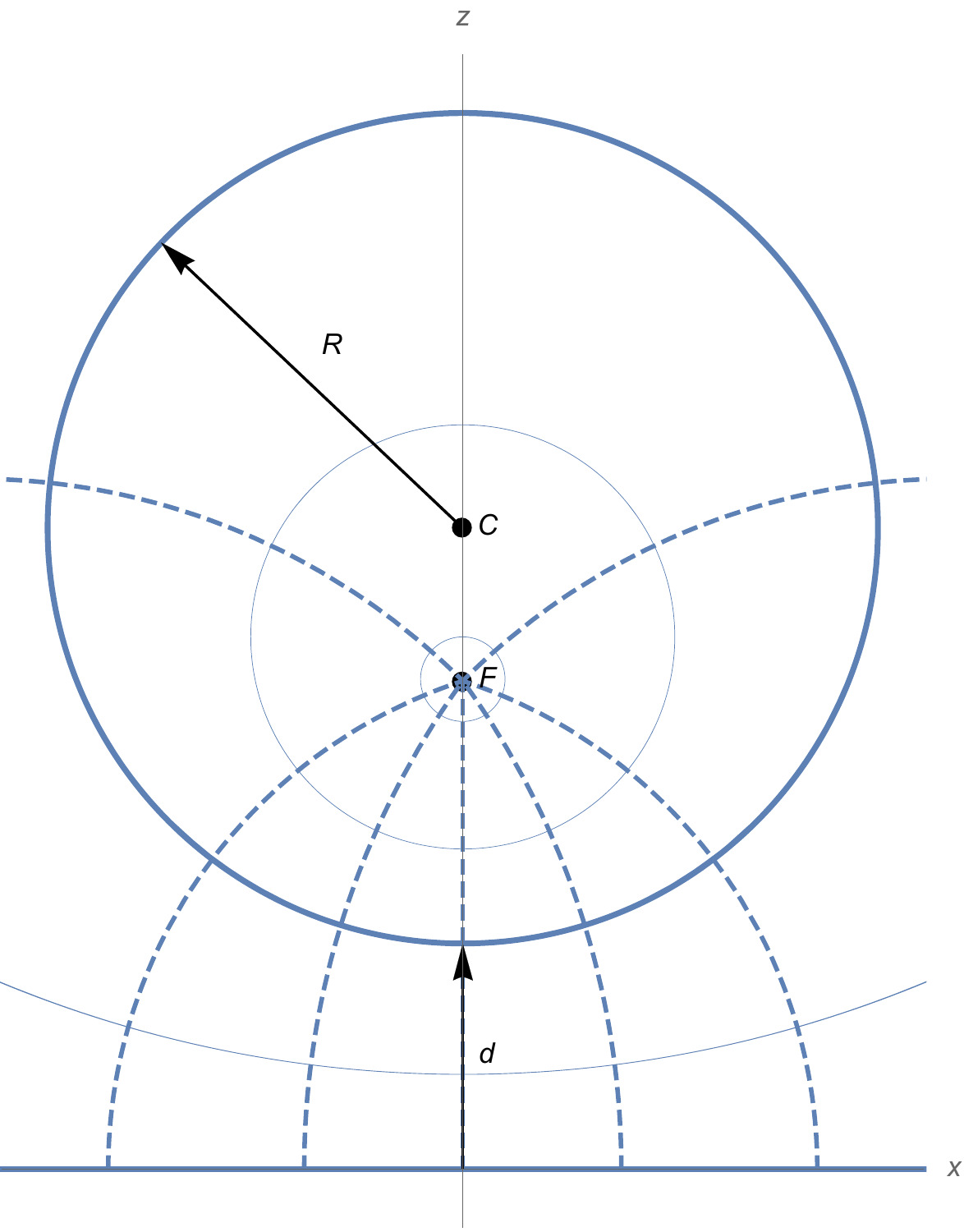}
\caption{\label{fig1} Geometry   of a sphere and plate. Shown are the center $C$ of the sphere and its focus $F$. The system is characterized by its aspect ratio $x=d/R$.  The thin solid and dashed  lines  correspond to curves of  constant bispherical coordinates  $\mu$ and $\eta$ respectively.}
\end{figure}
According to Eq. (\ref{HTscat}) the computation of ${\cal F}_{\rm HT}$ involves scattering of static em fields by  the two surfaces.   Static em fields with  $E$ and $M$  polarizations  represent, respectively, electrostatic and magnetostatic fields which do not mix under scattering by a dielectric surface of any shape. Therefore modes with $E$ and $M$ polarizations give separate contributions to the Casimir energy ${\cal F}_{\rm HT}$.   Moreover, it is easily seen that in the static limit the em scattering problem   is mathematically equivalent to the scattering problem for a    free scalar field obeying the Laplace Equation. For  perfect conductors, the bc obeyed by the scalar field are  as follows.  For $E$ polarization, the scalar field is subjected to either D or Dr bc on the surfaces of the two bodies,  depending on whether the plates are grounded or not \cite{bimonteex1,foscoDr2,foscoDr}, while for  $M$ polarization the scalar field obeys N bc.   The dimensionless function $\Phi^{\rm (P)}$ providing the    classical
Casimir interaction for P bc can be thus decomposed as the sum of two independent contributions  $\Phi^{\rm (D/Dr)}$ and $\Phi^{\rm (N)}$, corresponding respectively to a D/Dr and a N scalar field:
\be
\Phi^{\rm (P)}=\Phi^{\rm (D/Dr)}+\Phi^{\rm (N)}\;.\label{split}
\ee
In the limit of vanishing separations $x$, the Casimir energy approaches the PFA limit:
\be
\Phi^{\rm (D)}_{\rm PFA}=\Phi^{\rm (Dr)}_{\rm PFA}=\Phi^{\rm (N)}_{\rm PFA}=\frac{\Phi^{\rm (P)}_{\rm PFA}}{2}=\frac{\zeta(3)}{8 \, x}\;.
\ee
The {\it exact} expression of the functions $\Phi^{\rm (D/Dr)}$ was determined in \cite{bimonteex1} by taking advantage of the separability of Laplace Equation in bisperical coordinates \cite{morse}:
\be
\Phi^{\rm (D)}=-\frac{1}{2} \sum_{l=0}^{\infty} (2 l +1) \ln[1-Z^{2 l+1}]\;,\label{enerD}
\ee
$$
\Phi^{\rm (Dr)}=-\frac{1}{2} \left\{\sum_{l=1}^{\infty} (2 l +1) \ln[1-Z^{2 l+1}]\;\right.
$$
\be
\left.+\ln \left[1-(1-Z^2) \sum_{l=1}^{\infty} Z^{2 l+1} \frac{1-Z^{2 l}}{1-Z^{2 l+1}} \right] \right\}\;,\label{enerDr}
\ee
where the parameter $Z$ depends on the aspect ratio  $x$:
\be
Z=[1+x+\sqrt{x\,(2+x)}]^{-1}\;.\label{Zdef}
\ee
The parameter $Z$ is less than one for all positive values of $x$, and as $x$ increases from 0 to $\infty$, $Z$ decreases monotonically from 1 towards zero.

Unfortunately, for N bc the Casimir interaction  $\Phi^{\rm (N)}$ cannot be computed exactly. We find convenient to introduce the {\it difference} $\delta \Phi$ between the N and D energies $\Phi^{(N)}$ and   $\Phi_m^{(D)}$:
\be
\delta \Phi=\Phi^{(N)}-\Phi^{(D)}\;. \label{deltaphi}
\ee
The energy for ungrounded perfect-conductors is accordingly represented as:
\be
\Phi^{\rm (P)}=\Phi^{\rm (Dr)}+\Phi^{\rm (D)}+\delta \Phi\;,\label{split2}
\ee
while for grounded conductors we write:
\be
\Phi^{\rm (P)}|_{\rm gr}= 2 \,\Phi^{\rm (D)}+\delta \Phi\;.
\ee
In the next Section we shall work out an asymptotic formula for $\delta \Phi$, valid in the limit of small separations, while in Sec.  IV $\delta \Phi$ shall be computed numerically using the exact scattering formula Eq. (\ref{scafor}).

\section{A short-distance formula for $\delta \Phi$}

Before we start the computation of $\delta \Phi$, it is important to notice that, due to  the presence of the trace in the general scattering formula Eq. (\ref{scafor}),  the Casimir interaction depends only on the equivalence class $[[M]]$ formed by all matrices $M$  that represent the operator ${\hat M}$, where  two elements $M$ and $M'$ of $[[M]]$ differ  by a  similarity transformation by an invertible matrix $A$: $M'=A M A^{-1}$.

The matrix $M^{(\rm N)}$ for N bc is easily computed in a {\it spherical} multipole basis with origin at the sphere center $C$. In this basis   the regular and outgoing eigenfunctions of the Laplace Equation  have   the familiar form $\phi_{lm}^{(\rm reg)}=r^l Y_{lm}(\theta,\phi)$, and $\phi_{lm}^{(\rm out)}=r^{-(l+1)} Y_{lm}(\theta,\phi)$, with $l \ge 0$, and $m=-l, \cdots,l$. By rotational symmetry around the azimuthal axis ${\hat z}$, the matrix $M^{(\rm N)}$ commutes with $J_z$ and hence is block diagonal. We let  $M^{(\rm N |m)}$ the block corresponding to the value $m$ of $J_z$.  One finds:
\be
{\hat M^{(\rm N|m)}}=\left[ \left[\frac{l}{l+1} \, \frac{(l+l')!}{(l+m)! (l'-m)!} \left(\frac{1}{2(1+x)} \right)^{l+l'+1}  \right] \right]\;,\label{Nmat}
\ee
with $l,l' \ge |m|$.
Apart from the factor $l/(l+1)$,  the matrix ${\hat M^{(\rm N|m)}}$ coincides with the corresponding matrix ${\hat M^{(\rm D|m)}}$  for D bc:
\be
{\hat M^{(\rm D|m)}}=\left[ \left[  \frac{(l+l')!}{(l+m)! (l'-m)!} \left(\frac{1}{2(1+x)} \right)^{l+l'+1}  \right] \right]\;,\label{Dmat}
\ee
Each block ${\hat M}^{(\rm N |m)}$  contributes separately to the Casimir energy, and we denote by  $\Phi_m^{(N)}$  the corresponding contribution to  $\Phi^{(N)}$. Of course, opposite values of $m$ give identical contributions to the Casimir energy, i.e.  $\Phi_m^{(N)}= \Phi_{-m}^{(N)}$. We can thus write $\Phi^{(N)}$ as:
\be
\Phi^{(N)}=2 \sum_{m \ge 0} \,\!\!' \Phi_m^{(N)}\;,
\ee
where the prime again denotes that the $m=0$ term is taken with a weight 1/2 and
\be
\Phi_m^{(N)}=-\frac{1}{2} {\rm Tr \ln} [1-{\hat M}^{(\rm N|m)}]\;.\label{HTscat2}
\ee

\subsection{ Contribution of the modes with $m=0$.  }

Luckily enough the contribution $\Phi_0^{(N)}$ of the $m=0$ modes can be computed exactly.
By a similarity transformation with the diagonal matrix $A_{ll'}=(l+1) \delta_{ll'}$ the matrix $ { M^{(\rm N | 0)}}$ in Eq. (\ref{Nmat}) is transformed to the matrix $ {\tilde M^{(\rm N | 0)}}$
\be
{\tilde M^{(\rm N|0)}}_{ll'}= \frac{l}{l'+1} \, \frac{(l+l')!}{l! \; l'!} \left(\frac{1}{2(1+x)} \right)^{l+l'+1}  \;,\label{simM0}
\ee
with $l,l' \ge 0 $. The first row of the matrix $ {\tilde M}^{(\rm N | 0)}$   is zero, while its $l$-th row with $l=1,2,\cdots$   is   identical to the $(l-1)$-th row of the matrix $M^{(\rm D|0)}$ in Eq. (\ref{Dmat})  with its first column  deleted: ${\tilde M}^{(\rm N|0)}_{l l'}=M^{(\rm D|0)}_{l-1,l'+1}$, $l=1,2,\cdots$, $l'=0,1,2,\cdots$.
By a second similarity transformation with the upper diagonal matrix ${\tilde A}(Z)$:
\be
{\tilde A}_{ll'}(Z)=Z^{l'-l} \frac{l'!}{(l'-l)!\; l!}\;,
\ee
with ${\tilde A}^{-1}(Z)={\tilde A}(-Z)$, the matrix  $ {\tilde M}^{(\rm N | 0)}$   is transformed into a lower triangular matrix $ {\bar M}^{(\rm N|0)}$,  with diagonal elements   equal to
$ {\bar M}^{(\rm N|0)}_{ll}= Z^{2 l+3}$, $l=0,1,2,\cdots$.  This implies at once:
\be
\Phi_0^{(N)}=-\frac{1}{2}\sum_{l \ge 0} \ln [1-Z^{2 l +3}]\;.\label{phi0Nex}
\ee
This result can be contrasted with the analogous formula for D bc \cite{bimonteex1}:
\be
\Phi_0^{(D)}=-\frac{1}{2}\sum_{l \ge 0} \ln [1-Z^{2 l +1}]\;.\label{phi0Dex}
\ee

\subsection{Contribution  of  modes with $m \neq 0$. }

Unfortunately, the contributions $\Phi_m^{(N)}$ of the modes with $m \neq 0$ cannot be computed exactly.  By using the technique of Refs. \cite{bordag1,bordag2,teo1} it is however possible to
prove a short-distance formula for $\Phi_m^{(N)}$, or more precisely for the difference  $\delta \Phi_m=\Phi_m^{(N)}-\Phi_m^{(D)}$.
We start by expanding the logarithm in Eq. (\ref{HTscat2}):
\be
\Phi_m^{(N/D)}= \frac{1}{2} \sum_{s=0}^{\infty} \frac{1}{s+1}\left(\prod_{i=0}^{s} \sum_{l_i=|m|}^{\infty}\right)  \prod_{i=0}^{s} { M}^{(\rm N/D|m)}_{l_i,l_{i+1}} \;,\label{expan1}
\ee
where $l_{s+1} \equiv l_0$.
Next, for $0< i \le s$ we perform  on the indices $l_i$ the shift: $l_i = l+l_i'$, where we set $l:=l_0$. For small separations $x  \ll1$, the Casimir energy is dominated \cite{bordag1,bordag2,teo1} by multipoles such that:
\be
l \sim 1/x\;, \;\;\;\;|l_i'| \sim 1/ \sqrt{x}\;,\;\;\;\;|m| \sim 1/\sqrt{x}\;.\label{order}
\ee
For small $x$ the discrete sums over $l$ and $l_i'$ in Eq. (\ref{expan1}) can be replaced by    integrations  (this corresponds to taking the leading term in the Abel-Plana summation formula):
$$
\Phi_m^{(N/D)}= \frac{1}{2} \int_0^{\infty} \!\!\!dl \; \left[ { M}^{(\rm N/D|m)}_{l,l} \right.
$$
\be
\left.+ \sum_{s=1}^{\infty} \frac{1}{s+1} \left(\prod_{i=1}^{s}  \int_{-\infty}^{\infty} \!\!\!d l_i' \right)  \prod_{i=0}^{s} { M}^{(\rm N/D|m)}_{l+l_i',l+l_{i+1}'} \right]\;,\label{expan2}
\ee
and we set $l_0'=l_{s+1}' \equiv 0$. In writing the above Equation, we considered that the integration over $l$ extending from $|m|$ to $\infty$ can be  replaced by an integration from zero to $\infty$ because, according to Eq. (\ref{order}), in the limit of small separations $m$ is negligibly small compared to $l$. We similarly replaced   the integration over $l_i'$ extending from $|m|-l$ to $\infty$ by an integration from $-\infty$ to $\infty$  because, compared to $l_i'$,  $(|m|-l)$  can be identified with $- \infty$.
Next, we observe that by virtue of Eq. (\ref{order}) the numbers $l+l', l \pm m$ are all large integers for small $x$ and therefore the factorials in Eqs. (\ref{Nmat}) and (\ref{Dmat}) can be computed using Stirling's formula:
\be
\ln n! =\left(n+\frac{1}{2}\right) \ln n -n +\frac{1}{2}\ln 2 \pi+\frac{1}{12 n}+\cdots
\ee
At this point, we  Taylor expand  the {\it difference}   $\delta { M}^{(\rm m)}_{l+l_i',l+l_{i+1}'}={ M}^{(\rm N|m)}_{l+l_i',l+l_{i+1}'}-{ M}^{(\rm D|m)}_{l+l_i',l+l_{i+1}'}$ among the matrices $M^{(N)}$ and $M^{(D)}$  in powers of $\sqrt{x}$  (powers of $\sqrt{x}$ are reckoned according to the estimates in  Eqs. (\ref{order})). Up to terms of order $x^{2}$ we find:
\begin{widetext}
\be
\delta { M}^{(\rm m)}_{l+l_i',l+l_{i+1}'}=\frac{1}{\sqrt{4 \pi l}} \left(\frac{l}{l+1}-1 \right)\,\exp \left[-2 x l -\frac{(l_i'-l_{i+1}')^2}{4 l}-\frac{m^2}{l}\right]+o(x^2)\;.\label{deltaM}
\ee
 On the other hand,  by taking the Taylor expansion of Eq. (\ref{Dmat}) we find:
\be
 { M}^{(\rm D| m)}_{l+l_i',l+l_{i+1}'}=\frac{1}{\sqrt{4 \pi l}} \exp{\left[-2 x l -\frac{(l_i'-l_{i+1}')^2}{4 l}-\frac{m^2}{l}\right]}+ o(x)\;.\label{TayMD}
\ee
\end{widetext}

The two formulae above confirm correctness of  the estimates in Eq. (\ref{order}).
There is a tricky but important point to stress here:  following the logic of the Taylor expansion, one might  find appropriate to replace the factor $[l/(l+1)-1]$ in the r.h.s. of Eq. (\ref{deltaM}) by its first order Taylor approximation $[l/(l+1)-1]=-1/l+o(x^2) $. The problem with this substitution  is that it leads to an infra-red divergence in the integral over $l$. To avoid this problem, we keep the  complete factor $[l/(l+1)-1]$ in Eq.(\ref{deltaM}).

Starting from Eq. (\ref{expan2}), and making use of  Eqs. (\ref{deltaM}) and (\ref{TayMD}), we obtain the following expression for $\delta \Phi_m$:
\begin{widetext}
$$
\delta \Phi_m=\frac{1}{2} \int_0^{\infty}
\frac{d l}{\sqrt{4 \pi l}}\left\{ \left(\frac{l}{l+1}-1 \right) \exp \left(-2 x l  -\frac{m^2}{l}\right) +
 \frac{1}{2} \sum_{s=1}^{\infty} \frac{1}{s+1}
  \left[\left(\frac{l}{l+1}\right)^{s+1}-1 \right] \right.
$$
\be
\left. \times
 \left(\prod_{i=1}^{s}  \int_{-\infty}^{\infty}  \frac{d l_i'}{\sqrt {4 \pi l}} \right) \prod_{i=0}^{s} \exp \left[-2\, x\, l -\frac{(l_i'-l_{i+1}')^2}{4 l}-\frac{m^2}{l}\right]\right\}+o(x) \;,\label{expan4}
\ee
Performing the gaussian integrals over $l_i'$, we then obtain the following estimate for $\delta \Phi_m$ accurate to order $x^{1/2}$:
\be
\delta \Phi_m^{(1/2)}=
 \frac{1}{2} \sum_{s=0}^{\infty} \frac{1}{(s+1)^{3/2}}  \int_0^{\infty}
\frac{d l}{\sqrt{4 \pi l }}  \left[\left(\frac{l}{l+1}\right)^{s+1}-1 \right] \exp\left[(s+1) \left(-2\, x\, l  -\frac{m^2}{l}\right)\right]
   \;.\label{expan5}
\ee
The sum over $s$ can be expressed in terms  the polylogarithm function ${\rm Li}_n(z)=\sum_{k=1}^{\infty} z^k/k^n$:
\be
\delta \Phi_m^{(1/2)}=
 \frac{1}{2}   \int_0^{\infty}
\frac{d l}{\sqrt{4 \pi l }}  \left\{ {\rm Li}_{3/2}\left[\frac{l}{l+1} \exp\left(-2\, x\, l  -\frac{m^2}{l}\right)\right]-{\rm Li}_{3/2}\left[ \exp\left(-2\, x\, l  -\frac{m^2}{l}\right)\right]\right\}
   \;.\label{expan6}
\ee
\end{widetext}
Combining the above formula with the exact expressions of $\Phi_0^{(N)}$ (Eq. (\ref{phi0Nex})) and $\Phi_0^{(D)}$ (Eq. (\ref{phi0Dex})) we obtain for $\delta \Phi$ the approximate small distance formula:
\be
\delta \Phi^{(0)}=-\frac{1}{2}\sum_{l \ge 0} \ln \left(\frac{1-Z^{2 l +3}}{1-Z^{2 l +1}}\right)+ 2 \sum_{m>0} \delta \Phi_m^{(1/2)}\;.\label{delphi12}
\ee
We expect that this formula for $\delta \Phi$ is accurate to order $x^0$. We shall later see that, despite the assumption $x \ll 1$ made in its derivation,  the above formula provides a very precise value of $\delta \Phi$ also for relatively large separations (see Fig. \ref{deltaphi}).

\subsection{ Expressions at small distances}

With exact expressions for the Casimir energies in the D and Dr models, one can compute explicitly the interaction in the limit of short distances $x \ll 1$. This limit corresponds to Z close to unity,  and one can compute the series in Eqs. (\ref{enerD}) and (\ref{enerDr}) using the Abel-Plana formula. We set $Z=\exp(-\mu)$, and then expand for small $\mu$, where $\mu=\ln [1+x+\sqrt{x\,(2+x)}]$. The resulting analytical  expressions for the Casimir interaction were worked out in \cite{bimonteex1}, and we reproduce them here for the convenience of the reader:
\be
\Phi^{(\rm D)}=\frac{\zeta(3)}{4 \mu^2}-\frac{1}{24} \ln \mu-\frac{1}{16}+\gamma_0'+\frac{7}{5760}\mu^2+o(\mu^4)\;,\label{Dshort}
\ee
\be
\Phi^{(\rm Dr)}=\Phi^{(\rm D)}- \frac{1}{2} \ln(\gamma_1-\ln \mu)-\frac{1}{12} \frac{\ln \mu-\gamma_2}{\ln \mu-\gamma_1}\mu^2+o(\mu^4)\;,\label{Drshort}
\ee
with $\gamma_0'=0.0874485$, $\gamma_1=1.270362$, $\gamma_2=1.35369$. We used $\mu$ as a variable for the expansion, for it provides a very accurate
 result also at larger distances. Both the D and Dr energies depend only on $\ln \mu$ and {\it even} powers of $\mu$. This implies that the energies depend only on $\ln x$ and
{\it integer} powers of $x$. In particular, the force for the D case, once expanded in $x$, is a Laurent series starting from $1/x^2$. However, for the Dr case there are  logarithmic terms
in the force as well.  The leading correction to PFA is the same $\ln \mu$ term for both models,  and its  coefficient is in agreement with the  DE. Interestingly, for practically
relevant separations the subleading double logarithmic term in Eq. (\ref{Drshort}) dominates over the leading logarithmic term, and therefore the D and Dr energies display rather
different behaviors.

To work out the {\it leading} correction to PFA of the N energy, we start from Eq. (\ref{delphi12}). It is convenient to use for  $\delta \Phi_m^{(1/2)}$ the expression in Eq. (\ref{expan5}). In the limit of vanishing separation, the sum over the angular index $m$ can be replaced by an
integration over $m$ extending from $-\infty$ to $\infty$.  Performing the straightforward gaussian integral we find:
$$
\delta \Phi_{\rm as}=
 \frac{1}{4} \sum_{s=0}^{\infty} \frac{1}{\!\!(s+1)^2}  \int_0^{\infty}
\!\!\!{d l}   \left[\left(\frac{l}{l+1}\right)^{s+1} \!\!\!-1 \right] e^{-2(s+1) x l}
$$
\be
=\frac{1}{4}\int_0^{\infty}
\!\!\!{d l}   \left[{\rm Li}_2 \left(\frac{l}{l+1} e^{-2 x l}\right) -{\rm Li}_2 \left(  e^{-2 x l}\right) \right]  \;.
\ee
We computed analytically the asymptotic expansion of the above formula for $x \rightarrow 0$ and found its leading term:
\be
\delta \Phi_{\rm as}=-\frac{1}{16} \ln^2 x + o(\ln x) \;.\label{expan7}
\ee
Since in the D model the leading correction to the PFA is a $\ln x$ term (see Eq. (\ref{Dshort})), the   leading  correction to the PFA for the N model coincides with the  $\ln ^2 x$ term  of $\delta \Phi$:
\be
\Phi^{(N)}= \frac{\zeta(3)}{8 \, x}- \frac{1}{16} \ln^2 x+ o(\ln x)\;.
\ee
Earlier we pointed out that  the leading correction to the PFA  for the Dr and the D model is the same $\ln x$ term.  It then follows from Eq. (\ref{split2}) that the  $\ln ^2 x$ term  of $\Phi^{(N)}$    represents also the leading  correction to the PFA for   P bc:
\be
\Phi^{(P)}= \frac{\zeta(3)}{4 \, x}- \frac{1}{16} \ln^2 x+ o(\ln x)\;.\label{asymexp}
\ee
Thus our  analytical results provide a rigorous proof of the $\ln^2 (x)$ form of the leading curvature correction to PFA, in accordance with indications obtained from the high-precision numerical data of \cite{antoine}.

\section{Numerical computation of  $\delta \Phi$}

The  HT limit of the (ungrounded)   sphere-plate Casimir energy with P bc    was computed in \cite{antoine} by a large-scale numerical computation of  the exact scattering formula Eq. (\ref{scafor})  using the standard spherical basis with origin at the sphere center $C$.  The computation in \cite{antoine}  included up to 5000 partial wave orders, which
allowed the authors of Ref.  \cite{antoine} to accurately estimate  the functions $\Phi^{\rm (P)}$ and $\Phi^{\rm (Dr)}$ for aspects ratios $x \ge 2 \times 10^{-3}$.

Earlier we saw that the classical Casimir energy for ungrounded perfect conductors is the sum of the energies for a   Dr scalar plus a N scalar (see Eq. (\ref{split})). The (normalized) energy $\Phi^{\rm (Dr)}$ can be   computed exactly in the sphere-plate geometry (see Eq.  (\ref{enerDr})), while for N bc an exact formula exists for $m=0$ modes. In the previous Section we derived an asymptotic  formula, Eq. (\ref{delphi12}), valid for small-distances, for the difference $\delta \Phi$ among the HT Casimir energies for N and D bc.    In this Section the energy-difference $\delta \Phi$ is computed numerically, using the exact scattering formula Eq. (\ref{scafor}).  As we shall see,   $\delta \Phi$ can be computed very efficiently by using   a basis of {\it bispherical} multipoles \cite{bimonteex1}.

Bispherical coordinates $(\mu, \eta, \phi)$ \cite{morse} are defined by $(x,y,z)=a (\sin \eta \cos \phi, \sin \eta \sin \phi, \sinh \mu)/(\cosh \mu-\cos \eta)$, where $a$ identifies the focus $F$ of the sphere defined by $\mu=\mu_1>0$ (see Fig.\ref{fig1}). The sphere has radius $R=a/\sinh \mu_1$, and $L= a \coth \mu_1$ is the  distance of  its center $C$ from the $\mu=0$ plane. The Laplace Equation is separable in bispherical coordinates, and  its regular and   outgoing eigenfunctions are:
\be
\phi_{lm}^{\rm reg/out}=\sqrt{ \cosh \mu-\cos \eta} \, Y_{l m}(\eta, \phi) \exp[\pm (l+1/2)\mu]\;,\label{bisph}
\ee
for $l \ge 0$, $m=-l, \cdots, l$. Relative to the sphere (plane)  outgoing and regular eigenfunctions correspond, respectively  to the upper (lower) and lower (upper) sign in the exponential. Scattering solutions can be expanded in these eigenfunctions. It is a simple matter to verify that in the bispherical basis of Eq. (\ref{bisph}) the translation matrix $U$ is diagonal with elements
$U_{lml'm'}= Z^{l +1/2} \delta_{ll'} \delta_{mm'}$. where $Z=\exp(- \mu_1)$. For D  bc the $T$-matrix for  both the plane and sphere are minus  the identity operator. Therefore, in the bispherical basis the $M^{\rm (D)}$ matrix for D bc is {\it diagonal} with elements  $M^{\rm (D)}_{lml'm'}= Z^{2 l +1} \delta_{ll'} \delta_{mm'}$, and thus evaluation of the scattering formula Eq. (\ref{scafor}) is straightforward yielding the  result quoted in Eq. (\ref{enerD}). The case of Dr bc is more elaborate, as one has to remove the contribution of monopoles from the $m=0$ block. Details can be found in \cite{bimonteex1}.   For N bc the   $T$-matrix of the $\mu=0$ plane is equal to the identity operator.  However, the $T$-matrix of the sphere is unfortunately non-diagonal. Of course, the $T$-matrix is still block diagonal with respect to the  angular index $m$, and it is convenient to decompose its blocks as $T^{(2|m)}=1+ \delta T^{(N|m)}$.
By an explicit computation in the bispherical basis, it is found that the matrix $\delta T^{(N|m)}$ satisfies the linear system
\be
B^{(m)}  \;\delta T^{(N|m)}= -2 \sinh \mu_1 \;1\;,\label{linT}
\ee
where $B^{(m)}  $ is the matrix of elements
$$
B^{(m)}_{ll'}=[(2l+1) \cosh \mu_1+\sinh \mu_1] \delta_{ll'}
$$
\be
-(l-m)\delta_{l,l'+1}-(l'+m) \delta_{l+1,l'}\;.
\ee
with $l,l' \ge |m|$. The linear system Eq. (\ref{linT}) cannot be solved analytically, but it can be easily solved numerically after truncation in the multipole order $l,l' < l_{\rm max}$.

At this point it would seem that nothing is really gained by using bispherical multipoles, because we  still face the problem of computing determinants of infinite-dimensional matrices, as we had to do anyhow in the standard base of spherical multipoles. Indeed the situation seems   even worse now, because earlier at least the matrix   $M^{(N|m)}$   had a simple expression (see Eq. (\ref{Nmat})), while now the matrix $\delta T^{(N|m)}$ has to be itself computed numerically by  solving an infinite-dimensional linear system.  This shortcoming of  bispherical coordinates  is however rewarded by the crucial advantage of a much faster rate of convergence with respect to the maximum value $l_{\rm max}$ of the multipole index $l$. To see this, consider the expression of the $M$ matrix for N bc in bispherical coordinates:
\be
M^{(N|m)}_{ll'}=Z^{2l+1}(\delta_{ll'}+\delta T^{(m)}_{ll'})\;,
\ee
with $l,l' \ge |m|$.
When this expression is substituted into the scattering formula Eq. (\ref{scafor}), it is easy to factor out the D contribution, and one ends up
with the following exact representation for   the energy-difference  $\delta \Phi$ defined in Eq. (\ref{deltaphi}):
\be
\delta \Phi=- \sum_{m\ge 0}\,\!\!' \,{\rm Tr \log}[1+V^{(m)} \delta T^{(N|m)}] \;,\label{delbisp}
\ee
where  $V^{(m)} $ is the diagonal matrix of elements:
\be
V^{(m)}_{ll'}=\frac{1}{1-e^{\mu_1(2 l+1)}} \delta_{ll'}\;.
\ee
The exponential in the denominator of $V^{(m)}_{ll'}$ shows that the multipoles contributing to $\delta \Phi$ are those with $l,l' \lesssim 1/\mu_1$.   For small $x$,  $\mu_1=-\ln Z=\ln [1+x+\sqrt{x\,(2+x)}]^{-1} \simeq \sqrt{2 x}$ and then we see that the order $l$  of the relevant partial waves  scales like $\sqrt{R/d}$, which is only the square root of the multipole order $l_{\rm max} \sim R/d$ (see Eq. (\ref{order})) needed in the  spherical basis.

To demonstrate the fast rate of convergence of the scattering formula in bispherical coordinates, we take as an example  $x=2 \times 10^{-3}$, which is the smallest  aspect ratio considered in \cite{antoine}. By including in the scattering formula 5000 partial (spherical) waves, the authors of \cite{antoine} computed  $\Phi^{(\rm P)}=146.812$ and $\Phi^{(\rm Dr)}=74.5962$. On the other hand, using the exact formula in Eq. (\ref{enerD}) we find  $\Phi^{(\rm D)}=75.2936$. From Eq. (\ref{split2}) we then get $\delta \Phi=-3.07737$. In Table I we quote the values of $\delta \Phi$ obtained from Eq. (\ref{delbisp}) with inclusion of bispherical multipoles of order $l \le l_{\rm max}$, for $l_{\rm max}=20, 40, 80, 120$. As it can be seen, $\delta \Phi$ converges quickly, and already with $l_{\rm max}=80$ the error on $\delta \Phi$ is as small as $1.2 \times 10^{-5}$. With $l_{\rm max}=120$ we reproduce the value computed in \cite{antoine} using 5000 partial waves. It is interesting to compare the numerical value of $\delta \Phi$ with the estimate provided by the asymptotic formula Eq. (\ref{delphi12}).  Evaluation of  Eq. (\ref{delphi12}) gives  $\delta \Phi^{(0)}=-3.068$, which differs from the numerical value of $\delta \Phi$ by less than 0.3 \%.
\begin{table}[h]
\begin{tabular}{l l l l l } \hline
${ l_{\rm max}}$ \;\;\;&  20  &  40 & \;\;\;80  & \;\;\;120 \\ \hline \hline
$\delta \Phi$\;\;\; & -2.92435 \;\;\;& -3.06243 \;\;\;& -3.07725\;\;\; & -3.07737 \\  \hline
\end{tabular}
\caption{ Numerical values of $\delta \Phi$ for aspect ratio $x=2 \times 10^{-3}$  obtained from the scattering formula  in bispherical coordinates Eq.(\ref{delbisp}) with inclusion of multipoles of order $l \le l_{\rm lmax}$. The value of $\delta \Phi$ for $l_{\rm max}=120$ is in perfect agreement with the value obtained in  \cite{antoine} using spherical multipoles up to  $l_{\rm max}=5000$.}
\end{table}
\begin{figure}
\includegraphics [width=.9\columnwidth]{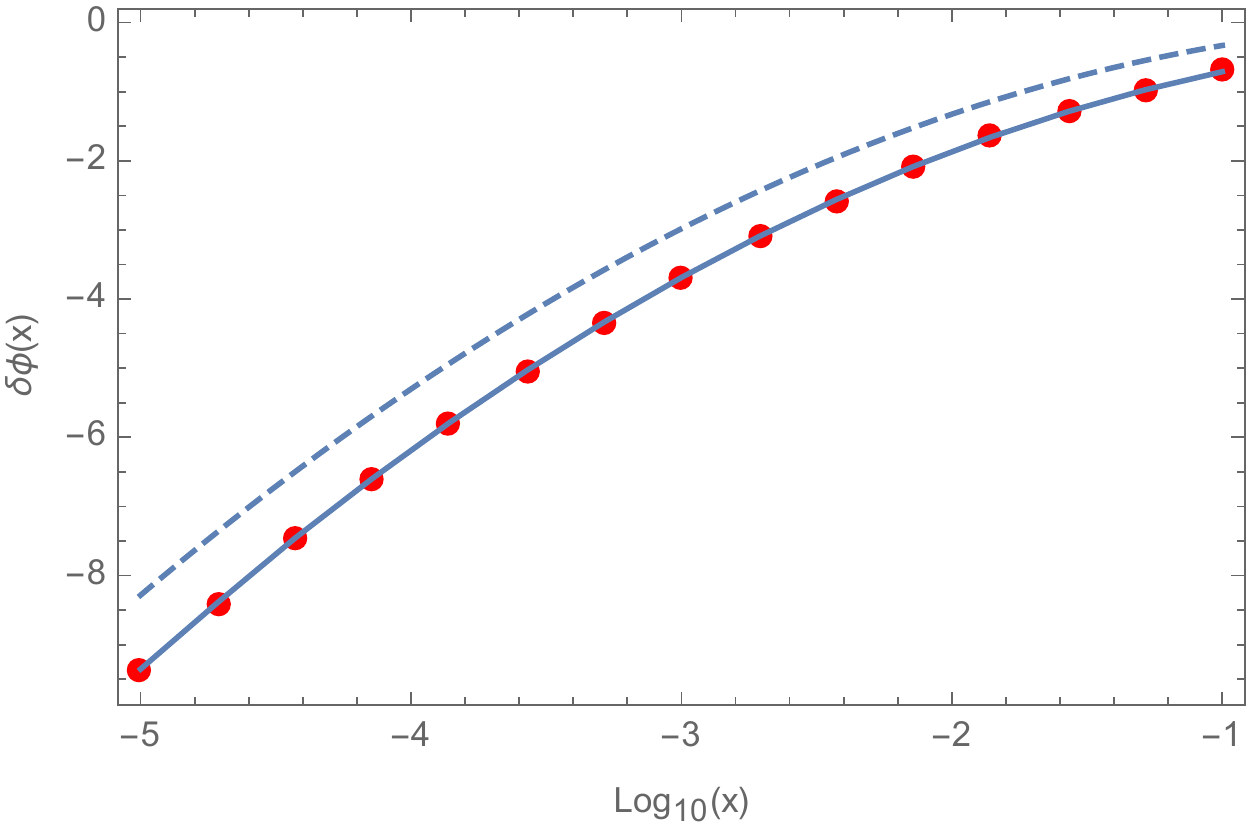}
\caption{\label{deltaphi}   Difference $\delta \Phi=\Phi^{(\rm N)}-\Phi^{(\rm D)}$ among the sphere/plate  N and D normalized Casimir energies as a function of $\log_{10}(x)$: numerical data (dots) computed using the scattering formula in a bispherical basis Eq. (\ref{delbisp}),   small-distance formula   Eq. (\ref{delphi12}) (solid line),   leading term Eq. (\ref{expan7}) (dashed line).}
\end{figure}

In Fig.\ref{deltaphi} we  plot   $\delta \Phi$ as a function $\log_{10}(x)$. The dots were computed using the scattering formula for $\delta \Phi$ in a bispherical basis Eq. (\ref{delbisp}). The fast convergence of   Eq. (\ref{delbisp}) allowed us to accurately compute $\delta \Phi$ for aspect ratios as small as $x=10^{-5}$ by using less than 1000 partial waves. The solid line Fig.\ref{deltaphi} was computed using the asymptotic formula for $\delta \Phi$, Eq. (\ref{delphi12}).   It can be seen   that the asymptotic formula Eq. (\ref{delphi12}) provides a precise estimate
of $\delta \Phi$  over the entire range of aspect ratios displayed in the Figure, up to the fairly large value $x=0.1$. The error made by using  Eq. (\ref{delphi12}) varies  from 0.16 \% for $x=10^{-5}$ to a maximum of 1.2 \% for $x=0.1$. The dashed line of Fig.\ref{deltaphi} corresponds to the leading term Eq. (\ref{expan7}). By a fit procedure, we verified that a very good agreement between the dashed curve   and the numerical data in Fig. \ref{deltaphi} can be obtained by adding to the expansion in Eq. (\ref{asymexp}) a subleading logarithmic term proportional to $\ln x$. 

\begin{figure}
\includegraphics [width=.9\columnwidth]{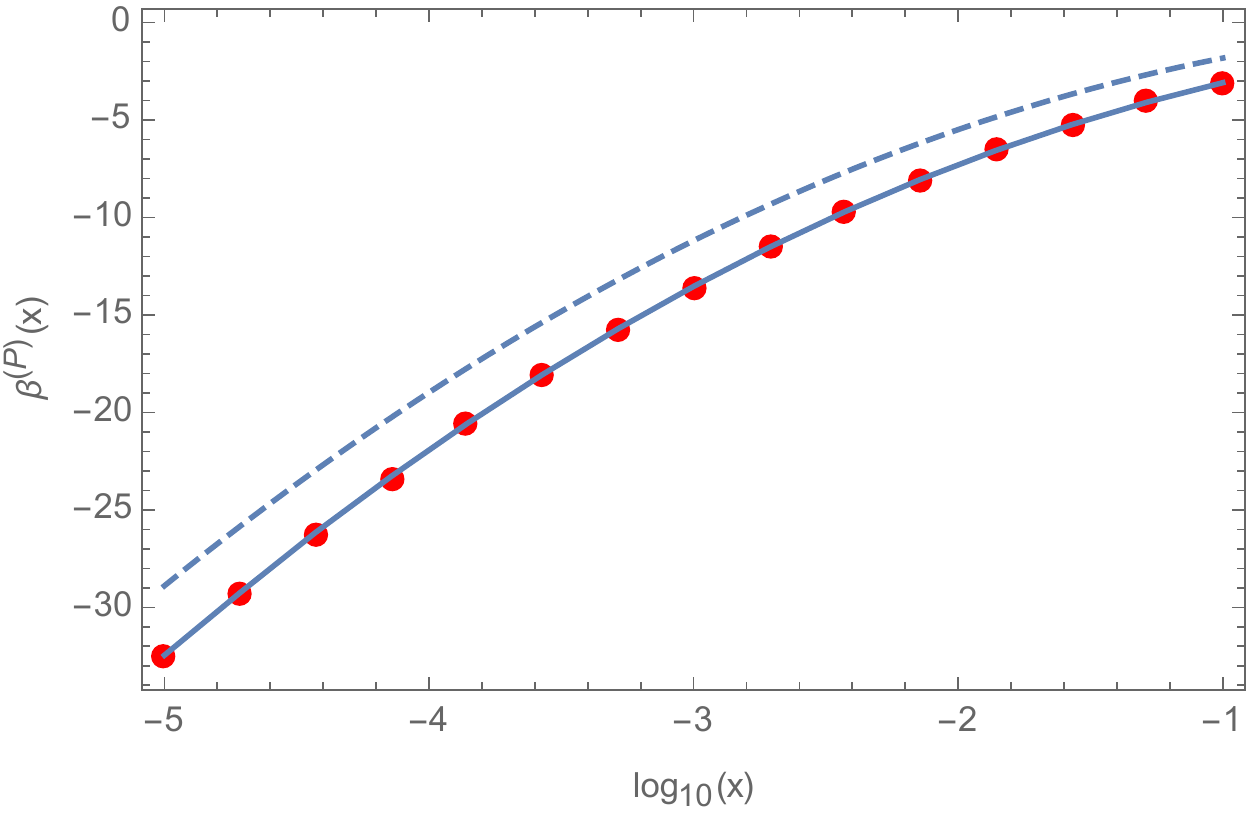}
\caption{\label{betaperf}  Additive correction $\beta^{\rm (P)}$ to the PFA classical Casimir energy for perfectly conducting sphere-plate as a function of  $\log_{10}(x)$: numerical data (dots) computed using the scattering formula   in a bispherical basis Eq. (\ref{delbisp}),   small-distance formula for $\delta \Phi$ Eq. (\ref{delphi12})  (solid line), leading term for $\delta \Phi$ Eq. (\ref{expan7}) (dashed line).}
\end{figure}

A convenient representation of deviations from the PFA energy is provided by the function $\beta^{\rm (P)}(x)$ defined such that \cite{antoine}:
\be
\Phi^{\rm (P)}=-k_B T \,\frac{ \zeta(3)}{4 } \left(\frac{1}{x}+ \beta^{(\rm P)}(x) \right)\;.
\ee
We similarly set:
\be
\Phi^{\rm (D/Dr/N)}=-k_B T \,\frac{ \zeta(3)}{8 } \left(\frac{1}{x}+ \beta^{(\rm D/Dr/N)}(x) \right)\;.
\ee
The exact expressions for the functions $\beta^{(\rm D/Dr)}(x)$  can be easily worked out starting from the exact solutions for the energies Eq. (\ref{enerD}) and (\ref{enerDr}).
On the other hand, $\beta^{(\rm N)}(x)$ can be expressed in terms of $\delta \Phi$:
\be
\beta^{(\rm N)}(x)=\beta^{(\rm D)}(x)+\frac{8}{\zeta(3)}\,\delta \Phi\;.
\ee
Recalling Eq. (\ref{split2}), $\beta^{(\rm P)}(x)$ can be decomposed as:
\be
\beta^{(\rm P)}(x)=\frac{1}{2}\left(\beta^{(\rm Dr)}(x)+\beta^{(\rm D)}(x)+ \frac{8}{\zeta(3)}\,\delta \Phi\right).
\ee
In Fig. \ref{betaperf} we display a plot of  $\beta^{\rm (P)}(x)$, where dots represent our numerical data, while the solid line is  computed using the   small-distance formula Eq. (\ref{delphi12}) for $\delta \Phi$.

\section{Conclusions}

We studied the Casimir interaction between a sphere and a plate, both perfectly conducting, in the classical limit of high temperatures. We worked out an analytical formula for the energy, valid for sufficiently small separations. Taking the asymptotic expansion of the small distance formula we  found a $\ln^2(d/R)$ correction in the energy, beyond the commonly used proximity force approximation. The $\ln^2(d/R)$ form of the correction is in agreement with a fit of  large-scale numerical data \cite{antoine}.  We developed a fast-converging numerical scheme for computing the Casimir energy, based on a system of bispherical partial waves. In bispherical coordinates, convergence  of the exact scattering formula is achieved  at multipole  order $l_{\rm max} \simeq \sqrt{R/d}$,  while in the standard approach based on spherical multipoles convergence is achieved only at order $l_{\rm max} \simeq R/d$.  Using the bispherical basis, we could accurately compute the Casimir energy for very small aspects ratio $d/R=10^{-5}$. Comparison with the high precision numerical data shows that the analytical small-distance formula  precisely estimates the energy also for fairly large values of the aspect ratio.

\acknowledgments

The author thanks T. Emig, N. Graham, M. Kruger, R. L. Jaffe and M. Kardar for valuable discussions while the manuscript was in preparation. Warm thanks are due also to the authors of \cite{antoine} for sharing with the author their numerical data.

\end{document}